\centerline{\bf The String Uncertainty Relations follow  }
\centerline{\bf from the New Relativity Principle }
\bigskip
\centerline{Carlos Castro}
\centerline{Center for Theoretical Studies of Physical Systems}
\centerline{Clark Atlanta University}
\centerline{Atlanta, GA. 30314}
\smallskip
\centerline{January 2000}

\bigskip

\centerline{\bf Abstract}

The String Uncertainty Relations have been known for some time
as the stringy corrections to the original Heisenberg's Uncertainty 
principle.
In this letter the Stringy Uncertainty relations,
and corrections thereof, are explicitly derived from the New Relativity
Principle
that treats all dimensions and signatures on the same footing and which 
is based on
the postulate that the Planck scale is the minimal length in Nature in 
the same vein
that the speed of light was taken as the maximum velocity in Einstein's 
theory of
Special Relativity. The Regge behaviour of the string's spectrum is also 
a natural
consequence of this New Relativity Principle.

\bigskip

Recently we have proposed that a New Relativity principle may be 
operating in Nature
which could reveal important clues to find the origins of $M$ theory  
[1]. 
We were forced to introduce this new Relativity principle, where all 
dimensions and
signatures of spacetime are on the same footing, to find a fully 
covariant formulation
of the $p$-brane Quantum Mechanical Loop Wave equations. This New 
Relativity Principle,
or the principle of Polydimensional Covariance as has been called by 
Pezzaglia,
has also been crucial in the derivation of Papapetrou's equations of 
motion of a
spinning particle in curved spaces that was a long standing problem
which lasted almost 50 years  [2]. A Clifford calculus was used where 
all the
equations were written in terms of Clifford-valued multivector 
quantities;
i.e one had to abandon the use of vectors and tensors and replace them 
by
Clifford-algebra valued quantities, matrices, for example .

In this letter we will explicitly derive the String Uncertainty 
Relations,
and corrections thereof, directly from the Quantum Mechanical Wave 
equations
on Noncommutative Clifford manifolds or {\bf C}-spaces  [1]. 
There was a one-to-one correspondence between the nested hierarchy of 
point, loop, {\bf 2}-loop,
{\bf 3}-loop,......{\bf p}-loop histories encoded in terms of 
hypermatrices
and wave equations written in terms of Clifford-algebra valued 
multivector quantities.
This permits us to recast the QM wave equations associated with the 
hierarchy of nested 
{\bf p}-loop histories, embedded in a target spacetime of $D$ dimensions 
, 
where the values of $p$ range  from :  $p=0,1,2,3......D-1$, as a 
$single$ QM line
functional wave equation whose lines live in a Noncommutative Clifford 
manifold of
$2^D$ dimensions. $p=D-1$ is the the maximum value of $p$ that 
saturates the
embedding spacetime dimension.

The line functional wave equation in the Clifford manifold, {\bf 
C}-space  is :

$$\int d\Sigma~( {\delta^2 \over \delta X(\Sigma) \delta X(\Sigma) } 
+{\cal E}^2 )
\Psi [X(\Sigma)]=0. \eqno (1)$$
where $\Sigma$ is an invariant evolution parameter of $l^{D}$ dimensions 
generalizing the notion of the invariant proper time in Special 
Relativity
linked to a massive point particle line ( path ) history :

$$(d\Sigma)^2 = (d\Omega_{p+1})^2 + \Lambda^{2p}(dx^\mu dx_\mu)
+  \Lambda^{2(p-1)}(d\sigma ^{\mu\nu}  d\sigma_{\mu\nu} )
+  \Lambda^{2(p-2)}(d\sigma^{\mu\nu\rho} d \sigma _{\mu\nu\rho})
+ .......\eqno (2)$$
$\Lambda$ is the Planck scale in $D$ dimensions. 
{\bf X}$(\Sigma)$  is a Clifford-algebra valued " line " living in the 
Clifford
manifold ( {\bf C}-space)  :

$$X=\Omega_{p+1} +\Lambda^p x_\mu \gamma^\mu +\Lambda^{p-1} \sigma_{\mu\nu} \gamma^\mu 
\gamma^\nu +\Lambda^{p-2}
\sigma_{\mu\nu\rho}\gamma^\mu \gamma^\nu \gamma^\rho +......... \eqno 
(3a)$$

The multivector {\bf X} encodes in one single stroke the point history 
represented by the
ordinary $x_\mu$ coordinates and the holographic projections of the 
nested family of   
{\bf 1}-loop, {\bf 2}-loop, {\bf 3}-loop...{\bf p}-loop histories onto 
the embedding
coordinate spacetime planes given respectively by : 
$$\sigma_{\mu \nu},  \sigma_{\mu \nu\rho}......\sigma_{\mu_1 
\mu_2...\mu_{p+1}}\eqno (3b)$$
The scalar $\Omega_{p+1}$ is the invariant proper 
$p+1=D$-volume associated
with the motion of the ( maximal dimension ) {\bf p}-loop across the 
$D=p+1$-dim target
spacetime. 
There was a coincidence condition [1] that required to equate the values 
of the
center of mass coordinates $x_\mu$, for all the {\bf p }-loops, with the 
values of the
$x^\mu$ coordinates of the
point particle path history. This was due to the fact that
upon setting $\Lambda=0$ all the {\bf p}-loop histories collapse to a 
point history. 
The latter history is the baseline where one constructs the whole 
hierarchy. 
This also required a proportionality relationship :

$$\tau \sim { A\over \Lambda }\sim {V \over \Lambda^2}\sim.......
\sim {\Omega^{p+1} \over \Lambda^p}. \eqno (4)$$
$\tau,A,V....\Omega^{p+1}$ represent the invariant proper time, proper 
area, proper volume,...
proper $p+1$-dim volume swept  by the  point, loop, {\bf 2}-loop, 
{\bf 3}-loop,.....
{\bf p}-loop histories across their motion through the embedding spacetime, respectively.
${\cal E}=T $ is a quantity of dimension $(mass)^{p+1}$, the maximal 
$p$-brane tension ( $p=D-1$) .

The wave functional $\Psi$ is in general a Clifford-valued, hypercomplex 
number.
In particular it could be a complex, quaternionic or octonionic valued 
quantity.
At the moment we shall not dwell on the very subtle complications and 
battles associated
with the quaternionic/octonionic extensions of Quantum Mechanics [14] 
based on Division algebras and simply take the wave function to be a 
complex number. 
The line functional wave equation for lines living in the Clifford 
manifold ( {\bf C}-spaces)
are difficult to solve in general. To obtain the String Uncertainty 
Relations, 
and corrections thereof,  one needs to simplify them.
The most simple expression is to write the simplified wave equation in units 
$\hbar=c=1$ :

$$[- ( {\partial ^2 \over \partial  x^\mu  \partial  x_\mu }
+{\Lambda^2\over 2}  {\partial ^2 \over \partial \sigma^{\mu\nu} 
\partial  \sigma_{\mu\nu}}
+ {\Lambda^4 \over 3!} {\partial ^2 \over \partial \sigma^{\mu\nu\rho} 
\partial
\sigma_{\mu\nu\rho} }
+......) - \Lambda^{2p} {\cal E} ^2 ]~\Psi [x^\mu, \sigma^{\mu\nu}, \sigma^{\mu\nu\rho},.....  ]=0 \eqno 
(5)$$
where we have dropped the first component of the Clifford multivector
dependence, $\Omega^{p+1}$, of the wave functional $\Psi$ and we have replaced 
functional differential
equations for ordinary differential equations. 
Had one kept the first component dependence   $\Omega^{p+1}$ on $\Psi$ one would have had a cosmological constant contribution to the 
${\cal E}$ term as we will see below.  Similar types of 
equations in a
different context  with only the first two 
terms of eq-(5),
have also been written in [2].

The last equation contains the seeds of the String Uncertainty Relations 
and corrections thereof.
Plane wave type solutions to eq-(5) are :

$$\Psi =e^{i ( k_\mu x^\mu + k_{\mu\nu}  \sigma^{\mu \nu} +  k_{\mu\nu\rho}  \sigma^{\mu \nu\rho}+ .......)}. \eqno (6)$$
where $k_{\mu\nu}, k_{\mu\nu\rho}.....$ are the area-momentum, volume-momentum,..... $p+1$-volume-momentum conjugate variables to the holographic 
$\sigma^{\mu\nu}, \sigma^{\mu\nu\rho}...$ coordinates respectively. These are the components of the Clifford-algebra valued 
$multivector$ {\bf K} that admits an expansion into a family of antisymmetric tensors of arbitrary rank like the Clifford-algebra valued "line" 
{\bf X} did earlier in eq-(3a). The multivector {\bf K} is nothing but the conjugate $polymomentum$  variable to {\bf X} in {\bf C}-space. 
Inserting the plane wave solution into the simplified wave equation  
yields the generalized 
dispersion relation, after reinserting the suitable powers of $\hbar$ :

$$\hbar^2  (k^2 +{1\over 2} \Lambda^2 (k_{\mu\nu} ) (k^{\mu\nu}) 
+{1\over 3!} \Lambda^4 (k_{\mu\nu\rho}  ) (k^{\mu\nu\rho} )+
........) - { \Lambda^{2p} {\cal E}^2 \over \hbar^{2p} }       =  0 . \eqno (7)$$
this is just the generalization of the ordinary wave/particle dispersion 
relationship

$$p^2 =\hbar^2 k^2 .~~~p^2-m^2=0 . \eqno (8)$$
Had one included the  $\Omega^{p+1}$ dependence on $\Psi$; i.e an extra piece $exp~[i\Omega_{p+1}\lambda ]$, where $\lambda$ is the cosmological constant 
of dimensions $(mass)^{(p+1)} $.  
The required $- \Lambda^{2p} \partial^2 \Psi / (\partial \Omega_{p+1})^2  $ 
term of the simplified wave equation (5) would have generated an extra term of the form $ \Lambda^{2p}\lambda^2 $. After reinserting the suitable powers of $\hbar$, 
the cosmological constant term will precisely $shift$ the value of the 
$- \Lambda^{2p} {\cal E}^2/\hbar^{2p} $ piece of eq-(7) to the value : $- ({ \Lambda \over \hbar} )^{2p}({\cal E}^2 - \lambda^2)$, which precisely has an overall  dimension of 
$m^2$ as expected.  

Hence, this will be then the " vacuum " 
contribution to $maximal$  $p$-brane tension ( $p=D-1$) : ${\cal E} =T_p$ has overall units $(mass)^{p+1}$; i.e energy per $p$-dimensional volume.         
On dimensional grounds and due to the $coincidence$ condition [1] referred above one has that :

$$(k_{\mu\nu}    ) (k^{\mu\nu}  )\sim (k^2)^2 =k^4.
~~~(k_{\mu\nu\rho}  ) (k^{\mu\nu\rho})\sim (k^3)^2 =k^6......\eqno (9)$$
where the proportionality factors in eq-(9) are $scalar$-valued quantities that we choose to be ( for simplicity ) 
the dimension-dependent constants, $\beta_1, \beta_2....$ respectively. 
The coincidence condition implies that upon setting $\Lambda =0$ all the {\bf p}-loop histories $collapse$ to a point history. 
In that case the areas, volumes, ...hypervolumes 
collapse to $zero$ and the wave equation (5) reduces to the ordinary Klein-Gordon equation for a spin zero massive particle.

Factoring out the  $k^2$ factor in (7), using the analog of the 
dispersion relation (8) and taking the square root, after performing the
binomial/Taylor expansion of the square root, 
subject to the condition $\Lambda^2  k^2 << 1$, 
one obtains an $effective$ energy dependent Planck " constant " that
takes into account the Noncommutative nature of the Clifford manifold 
({\bf C}-space )
at Planck scales :

$$\hbar_{eff} (k^2)  =\hbar (1 +{1\over 2.2!} \beta_1 \Lambda^2 k^2 +
{1\over 2.3!} \beta_2 \Lambda^4  k^4
+...................). \eqno (10) $$
where we have included explicitly the $D$ dependent coefficients
$\beta_1, \beta_2,...$ that arise in (9) due to the $coincidence$ condition and on dimensional analysis.

Arguments concerning an effective value of Planck's " constant " related 
to higher
derivative theories and the modified uncertainty relations have been 
given by [8]. 
The advantage of this derivation based on the New  Relativity Principle 
is that one
automatically avoids the problems involving the ad hoc introduction of 
higher derivatives in
Physics ( ghosts, ...) .

The uncertainty relations for the coordinates-momenta follow from the
Heisenberg-Weyl algebraic relation familiar in QM :

$$\Delta x \Delta p \ge | <[{\hat x} , {\hat p} ]> |. ~~~
[{\hat x}  , {\hat p} ] =i\hbar  \eqno (11)$$
Now we have that in {\bf C}-spaces, $x,p$ must not, and should not, 
be interpreted as ordinary vectors of spacetime but as one of the many
$components$ of the Clifford-algebra valued multivectors 
that " coordinatize "  the Noncommutative Clifford Manifold, {\bf 
C}-space.
The Noncommutativity is $encoded$ in the $effective$ value of the 
Planck's " constant "
which $modifies$  the Heisenberg-Weyl $x,p$ algebraic commutation 
relations and, consequently,
generates new uncertainty relations :

$$ \Delta x \Delta p \ge | <[ {\hat x} , {\hat p} ] > | = <\hbar_{eff}> =
\hbar (1 +{1\over 2.2!} \beta_1 \Lambda^2 <k^2> +
{1\over 2.3!} \beta_2 \Lambda^4 <k^4> +.......)    \eqno (12)$$

Using the relations :

$$\hbar k =p.~~~ <p^2 > \ge (\Delta p)^2. ~~~<p^4 > \ge (\Delta 
p)^4.....\eqno (13)$$
one arrives at :

$$\Delta x \Delta p \ge  \hbar + { \beta_1 \Lambda^2 \over 4 \hbar} 
(\Delta p)^2 +
{\beta_2 \Lambda ^4 \over 12 \hbar^3} (\Delta p)^4 +....... \eqno (14)$$

Finally, keeping the first two terms in the expansion in the r.h.s of 
eq- (14)
one recovers the ordinary String Uncertainty Relation [5] directly from 
the New Relativity Principle as promised :

$$\Delta x \ge {\hbar \over \Delta p} + {\beta_1 \Lambda^2 \over 4 
\hbar}(\Delta p).\eqno (15)$$
which is just a reflection of the minimum distance condition in Nature 
[3,4,5,6,7,10] 
and an inherent Noncommutative nature of the Clifford manifold  ( {\bf 
C}-space ). Eq-(15) yields a $minimum$ value of $\Delta x$ of the order of the Planck length $\Lambda$ 
that can be verified explicitly simply by minimizing eq-(15). 

There is a widespread misunderstanding
about the modification of the Heisenberg-Weyl algebra (12). One could 
start from a canonical
pair of variables $q,p$ and perform a $noncanonical$ change of variables 
$Q,P$ such as to
precisely reproduce the modified commutation relations of eq-(12) : 

$$ x\rightarrow x'=x. ~~p\rightarrow p'.~~[x', p']=
i\hbar [{\partial\over \partial p} , p' ] =i\hbar {\partial  p'  
\over \partial p}=i\hbar+ {i \beta_1 \Lambda^2 \over 4 \hbar} (p')^2 +.... \eqno (16a)$$
Integrating (16a) keeping only the leading terms yields the desired relationship between $p$ and $p'$  :

$$ p (p') =  \int {dp' \over  1 + {\beta_1 \Lambda^2 \over 4 \hbar^2 } (p')^2 } . \eqno (16b) $$ 

This $noncanonical$ change of coordinates is $not$ what is represented 
here by the modified
Heisenberg-Weyl algebra. Space at
small scales is $not$ necessarily governed by the familiar Lorentzian 
symmetries : it is a
Noncommutative Clifford manifold that requires abandoning the naive 
notion of vectors and tensors
and replacing them by Clifford multivectors. Inotherwords, it is a world 
where Quantum Groups
operate . The fact that the String Uncertainty relations reflect the 
existence of a
minimum length in Nature 
is consistent with the $discretization$ of spacetime at the Planck scale 
[9]
and the replacement of ordinary Lorentzian group symmetries by Quantum 
Group ( Hopf Algebras)
Symmetries [12].

The New Relativity Principle reshuffles, for example, 
a loop history into a membrane history; a membrane history into a
into a $5$-brane history; a $5$-brane history into a $9$-brane history  and so forth; in particular it can transform a $p$-brane history into suitable combinations 
of other $p$-brane histories as building blocks. This is the bootstrap idea taken from the point particle case to to the $p$-branes case : 
each brane is made out of all the others. " Lorentz" transformations in {\bf 
C}-spaces
involve hypermatrix changes of " coordinates " [1] . The naive Lorentz 
transformations do not
apply in the world of Planck scale physics. Only at large scales the 
Riemannian continuum is
recaptured . For a discussion of the more fundamental Finsler Geometries implementing the minimum scale ( maximal proper acceleration ) in String 
Theory see [13].

The New Relativity principle not only reproduces the ordinary String 
Uncertainty Relations but
yields $corrections$ thereof in one single stroke as we have shown in 
eq-(14)!
This is a positive sign that the New Relativity principle is on the 
right track to reveal the
geometrical foundations of $M$ theory .
Uncertainty relations based on the Scale Relativity theory [3] were 
furnished in [10].
We must emphasize that the latter uncertainty relations involved 
spacetime $resolutions$.
Resolutions are $not$ statistical uncertainties,
therefore the relations [10] $cannot$ be used
to evaluate the modified coordinates-momenta commutation relations like 
the r.h.s of (12).

To finalize this letter we will show how the Regge trajectories 
behaviour of the
string's spectrum emerges also from the New Relativity principle.
Pezzaglia's derivation of Papapetrou's equations [2] for a spinning 
particle
moving in curved spaces were based on an invariant interval of the form 
:

$$ (d\Sigma)^2 =dx^\mu dx_\mu + {1\over 2 \lambda^2} d\sigma^{\mu\nu} 
d\sigma_{\mu\nu}.
\eqno (17a)$$
where $\lambda$ is a length scale. The norm of the Clifford-valued 
momentum is :

$$P^2 =p_\mu p^\mu + {1\over 2 \lambda^2} S_{\mu\nu} S^{\mu\nu}. \eqno 
(17b)$$
where $S^{\mu\nu}$ is the spin or canonically-conjugate variable to
the $area$ . If we set the $\lambda \sim \Lambda$ and relate the 
squared-norm 
$||S^{\mu\nu}||^2 $ to the value of the norm-squared of the {\bf 
2}-vector
conjugate to the holographic area variables $\sigma^{\mu\nu}$ ; i.e
$(k^{\mu\nu} )(k_{\mu\nu} )$, norm-squared  which is
proportional to $k^4$, 
one can infer, after inserting the appropriate units ( $c=1$), from  
the spin-squared  terms of
eq-(17b) and the dispersion relation given by eq-(7) :

$${||S^{\mu \nu}|| \over \hbar} \sim {\Lambda^2 k^2 }= {\Lambda^2 
p^2\over  \hbar^2} =
{\Lambda^2 m^2 \over  \hbar^2 } =n ({\Lambda^2 m^2_P \over \hbar^2}) 
=n. \eqno (18)$$
hence one has that the spin is quantized in units of $\hbar$ and
from the third term in the r.h.s of eq-(18) one
recovers the Regge trajectories behaviour of the string spectrum in 
units where
$\hbar=c=1$ :

$$J \sim \alpha' m^2 +a.~~~ m^2 \sim n m^2_P.~~~\alpha' \sim \Lambda^2 
={1\over m^2_P}.
\eqno (19)$$ 
which is consistent with the action-angle variables/ area-quantization  $A=n\Lambda^2$ ( in units of $\hbar =c=1$) :

$$S=\int P_{\mu\nu}d\sigma^{\mu\nu}\sim T A \sim T (n \Lambda^2) \sim n {\Lambda^2  \over 2\pi \alpha '}  \sim n. \eqno (20)$$
where $P_{\mu\nu}$ is the area-momentum variable conjugate to the string 
areal interval
$d\sigma^{\mu\nu}$. It is the string analog of the ordinary momentum $p=mv$ for a point particle. 
The action is a multiple of $\hbar$ as the Bohr-Sommerfield action-angle relation indicates : $S =\oint pdg =n\hbar$. The area quantization 
condition, as well as the Bekenstein-Hawking entropy-area relation,  have  also been obtained by Loop Quantum Gravity methods [9]. Using Loop Quantum Gravity 
methods they arrive at $A \sim \Lambda^2 \sum \sqrt {j_i(j_i+1)} $ where one is summing over spin quantun numbers along the edges of a spin network. 
The New Relativity principle is telling us from eqs-(18,19,20) that $A=n\Lambda^2$ where $n$ is the spin quantum number !

Having derived the string uncertainty relations, and corrections 
thereof, and explained in
simple terms why there is a link between the 
Regge trajectories behaviour of the string spectrum with the quantization of area,  
should be enough
encouragement to proceed forward with the New Relativity Principle.

\smallskip

\centerline{\bf Acknowledgements}
\smallskip 

We thank Luis and Sheila Perelman for their hospitality in New York City 
where this work was completed, and for the use of their computer
facilities. We extend our gratitude to T. Riley for his gracious 
assistance. 
\smallskip  

\centerline{\bf References}

1. C. Castro : " Hints of a New Relativity Principle from $p$-brane 
Quantum Mechanics "

hep-th/9912113.

2. W. Pezzaglia : " Dimensionally Democratic Calculus and Principles of 
Polydimensional

Physics " gr-qc/9912025.

3. L. Nottale : Fractal Spacetime and Microphysics, Towards the Theory 
of Scale Relativity

World Scientific 1992.

L. Nottale : La Relativite dans Tous ses Etats. Hachette Literature. 
Paris. 1999.

4. M. El Naschie : Jour. Chaos, Solitons and Fractals {\bf vol 10} nos. 
2-3 (1999) 567.

5. D. Amati, M. Ciafaloni, G. Veneziano : Phys. Letts {\bf B 197} (1987) 
81.

D. Gross, P. Mende : Phys. Letts {\bf B 197} (1987) 129.

6. L. Garay : Int. Jour. Mod. Phys. {\bf A 10} (1995) 145.

7. A. Kempf, G. Mangano : " Minimal Length Uncertainty and Ultraviolet

Regularization " hep-th/9612084.

G. Amelino-Camelia, J. Lukierski, A. Nowicki : " $\kappa$ deformed 
covariant phase

space and Quantum Gravity Uncertainty Relations " hep-th/9706031.

8. R. Adler, D. Santiago : " On a generalization of Quantum Theory :

Is the Planck Constant Really Constant ?  " hep-th/9908073

9. C. Rovelli : " The century of the incomplete revolution...."

hep-th/9910131

10. C. Castro : Foundations of Physics Letts {\bf 10} (1997) 273.

11. A. Connes : Noncommutative Geometry. Academic Press. New York. 1994.

12. S. Mahid : Foundations of Quantum Group Theory. Cambridge University

Press. 1995. Int. Jour. Mod. Phys {\bf A 5} (1990) 4689.

L.C. Biedenharn, M. A. Lohe    :  Quantum Groups and q-Tensor Algebras . 
World

Scientific. Singapore . 1995.

13. H. Brandt : Jour. Chaos, Solitons and Fractals {\bf 10} nos 2-3 
(1999) 267.

14. S. Adler : Quaternionic Quantum Mechanics and Quantum Fields .

Oxford, New York. 1995.

\bye